\documentclass[aps,pra,prerprint,twocolumn,superscriptaddress,amsmath,amssymb,preprintnumbers,showpacs]{revtex4}

\usepackage{graphicx}% Include figure files
\usepackage{dcolumn}% Align table columns on decimal point

\begin{document}

%\draft

\title{Misbelief and misunderstandings on the non--Markovian dynamics of a damped harmonic oscillator}


\author{S. Maniscalco}
\affiliation{INFM, MIUR, and Dipartimento di Scienze Fisiche ed
Astronomiche dell'Universit\`{a} di Palermo, via Archirafi 36,
90123 Palermo, Italy.} \email{sabrina@fisica.unipa.it}

\author{F. Intravaia}
\affiliation{Laboratoire Kastler-Brossel, \'{E}cole Normale
Sup\'{e}rieure, CNRS, Universit\'{e} Pierre et Marie Curie 4,
place jussieu Case 74, Tour 12 F-75252, Cedex 05 Paris, France}

\author{J.Piilo}
\affiliation{Department of Physics, University of Turku, FIN-20014
Turun yliopisto, Finland}
\affiliation{Helsinki Institute of
Physics, PL 64, FIN-00014 Helsingin yliopisto, Finland}

\author{A. Messina}
\affiliation{INFM, MIUR and Dipartimento di Scienze Fisiche ed
Astronomiche dell'Universit\`{a} di Palermo, via Archirafi 36,
90123 Palermo, Italy.}

\vspace{1cm}

\begin{abstract}
We use the exact solution for the damped harmonic oscillator to
discuss some relevant aspects of its open dynamics often mislead
or misunderstood. We compare two different approximations both
referred to as Rotating Wave Approximation. Using a specific
example, we clarify some issues related to non--Markovian
dynamics, non--Lindblad type dynamics, and positivity of the
density matrix.
\end{abstract}


\maketitle

\section{Introduction}

The theory of open quantum systems deals with the dynamics of
quantum systems interacting with their surroundings. The most
common approach to the description of the time evolution of the
system stems from the fact that the total system, i.e.,
subsystem plus environment, is  closed. Hence the density matrix,
containing all the information on the state of the total system,
satisfies the Liouville-Von Neumann Master Equation. However, due
to the eventual complexity of the total system, this Master
Equation usually cannot be solved neither analytically nor
numerically. Moreover, the density matrix of the total
system often contains much more information than what we actually need,
since, by definition, we call \lq system\rq~that part of the
universe which is of interest for our study . In other words,  we focus our
attention only on the degrees of freedom of the system.
Mathematically this corresponds to a trace over the environmental
variables leading to the reduced density matrix of the system
$\rho_S$. The dynamics of $\rho_S$ can be rather involved because
of the effects of the interaction with the environment. In general,
understanding these effects  is not an easy task.

In this paper we concentrate on a paradigmatic model of
the theory of open systems, namely the damped harmonic oscillator.
We consider a single harmonic oscillator interacting with a
quantized environment modelled as an infinite chain of
non-interacting oscillators. This model, also known as quantum
Brownian motion (QBM) model,  is central in many physical
contexts, e.g., in quantum field theory \cite{cohen}, quantum optics
\cite{petruccionebook,carmichael,buzeck} and solid state physics
\cite{weiss}. The importance of the damped harmonic oscillator
model is also due to the fact that it is one of the few
non-trivial
systems for which an exact Master Equation for the reduced
density matrix can be formulated and exactly solved
\cite{Haake,Feynman,Caldeira,Hu,Ford,Grabert,PRAsolanalitica}. For
this reason it has been extensively studied both for fundamental
and for applicative research. On the one hand, indeed, it allows to
gain new insight in the process of decoherence, responsible for
the appearance of a classical world from the quantum world \cite{giulini}.
On the
other hand, it is the key model for many types of new quantum
technologies such as quantum computation \cite{steane}, quantum
cryptography \cite{crypt}, and
quantum teleportation \cite{telep}.

In this paper we discuss a recent analytic approach
\cite{PRAsolanalitica} aimed at solving the generalized Master
Equation describing the reduced system dynamics in very general
conditions. The method does not rely, indeed, neither on the
Born--Markov approximation, leading to a coarse grained
description of the dynamics, nor on the weak coupling assumption,
valid for quasi--closed system. We will give an expression of the
density matrix easily readable in physical terms since it is
simply related to diffusion and dissipation
coefficients. Moreover, we will use this analytic method to discuss
some points usually mislead. In particular two different
approximations both usually (and confusingly) referred to as
Rotating Wave Approximation (RWA) will be compared.
We give examples clarifying the relationship between the
following issues: non--Markovian dynamics, non--Lindblad type
dynamics, and the positivity of the density matrix.

The paper is organized as follows. In Sec.~I we introduce the
generalized Master Equation and its exact solution. In Sec.~II we
discuss the two types of RWAs and single out a  class of observables
not sensitive to the presence of rapidly oscillating terms. In
Sec.~III we study the short time non--Markovian dynamics of the mean
energy of the system and show the connection with Lindblad or
non--Lindblad type behavior. Finally, in Sec. IV conclusions are
presented.

\section{Exact dynamics}
\label{sec:I}

Let us consider a harmonic oscillator of frequency $\omega_0$
surrounded by a generic environment. The Hamiltonian $H$ of
the total system can be written as follows
\begin{equation}
H=H_0 + H_E + \alpha X E, \label{eq:secI1}
\end{equation}
where $H_0=\omega_0\left(P^2+X^2\right)/2$, $H_E$ and $\alpha X E$ are the
system, environment and interaction Hamiltonians, respectively,
and $\alpha$ is the dimensionless coupling constant. The
interaction Hamiltonian  considered here has a simple bilinear
form with  position of the oscillator $X$ and  position
environmental operator $E$. For the sake of simplicity we have
written the previous expressions in terms of dimensionless
position and momentum operators and set $\hbar=1$.
 We denote the
density matrix for the oscillator-environment system by $\rho$.

Under the  assumptions that: (i)  at $t=0$ system and environment
are uncorrelated, that is $\rho(0)= \rho_S (0) \otimes
\rho_E(0)$, with $ \rho_E$ the density matrix of the
environment; (ii) the environment is
stationary, that is $[H_E,\rho_E(0)]=0$; (iii) the expectation
value of the environmental operator $E$  is equal to zero, that is ${\rm
Tr_E} \left\{ E \rho_E(0) \right\}=0$ (as for example in the case
of a thermal reservoir), one can derive the following master
equation for QBM \cite{EPJRWA}
\begin{eqnarray}
\frac{d \rho_S(t)}{dt} = -i \mathbf{H}_{0}^S
 -  \Big[ \Delta(t) (\mathbf{X}^S)^2 - \Pi(t) \mathbf{X}^S
\mathbf{P}^S \nonumber \\
 - \frac{i}{2} r(t) (\mathbf{X}^2)^S + i \gamma(t) \mathbf{X}^S
\mathbf{P}^{\Sigma} \Big]\rho_S(t). \label{QBMme}
\end{eqnarray}

We indicate with $\mathbf{X}^{S(\Sigma)}$
and $\mathbf{P}^{S(\Sigma)}$ the commutator (anticommutator)
position and momentum operators respectively, and with
$\mathbf{H}_{0}^S$ the commutator superoperator relative to the
system Hamiltonian. Such a Master Equation, obtained by
using the time-convolutionless projection operator technique
\cite{annals,timeconv}, is the superoperatorial version of the
Hu-Paz-Zhang Master Equation \cite{Hu}. Let us note, first of all,
that the Master Equation (\ref{QBMme}) is local in time, even if
non-Markovian. This feature is typical of all the generalized
Master Equations derived by using the time-convolutionless
projection operator technique \cite{petruccionebook,Breuer99a} or
equivalent approaches such as the superoperatorial one presented
in \cite{PRAsolanalitica}.

The time dependent coefficients appearing in the Master Equation
are defined in terms of the noise and dissipation kernels
\cite{petruccionebook} and contain all the information about the
short time system-reservoir correlation. The coefficient $r(t)$
gives rise to a time dependent renormalization of the frequency of
the oscillator. In the weak coupling limit one can show that this
gives a negligible contribution as far as the reservoir cut-off
frequency remains finite \cite{petruccionebook}. The term
proportional to $\gamma(t)$ is a classical damping term while the
coefficients $\Delta(t)$ and $\Pi(t)$ are diffusive terms.

The superoperatorial Master Equation (\ref{QBMme}) can be exactly
solved by using specific algebraic properties of the
superoperators \cite{PRAsolanalitica}. The solution for the
density matrix of the system is derived in terms of the Quantum
Characteristic Function (QCF) $\chi_t(x,p)$ at time $t$, defined
by
\begin{equation}
\label{sdef} \rho_S(t)=\frac{1}{2\pi}\int \chi_t(x,p)\:
e^{-i\left(p X-x P\right)} dxdp.
\end{equation}
It is worth noting that one of the advantages of the
superoperatorial approach is the relative easiness in
calculating the analytic expression for the mean values of
observables of interest by using the relations:
\begin{eqnarray}
\langle X^n \rangle &=& (-i)^n\left(\frac{\partial^n}{\partial p^n
}\chi(x,p)\right)_{x,p=0},\nonumber \\
\label{xp}\langle P^n \rangle
&=&(i)^n\left(\frac{\partial^n}{\partial x^n
}\chi(x,p)\right)_{x,p=0}.
\end{eqnarray}

Neglecting the frequency renormalization terms, the exact analytic
expression for the time evolution of the QCF is
\cite{PRAsolanalitica}
\begin{equation}
\chi_t (x,p)=e^{- A_t(x,p)} \chi_0 \left[ e^{- \Gamma (t)}
\tilde{x}, e^{- \Gamma (t)} \tilde{p}  \right], \label{chit}
\end{equation}
with $\chi_0$ QCF of the initial state of the system and
\begin{eqnarray}
\tilde{x} &=& \cos(\omega_0 t) x + \sin(\omega_0 t) p, \nonumber \\
\tilde{p} &=& - \sin(\omega_0 t) x + \cos(\omega_0 t) p.
\end{eqnarray}
The quantity $A_t(x,p)$ is a quadratic form in the position $x$
and momentum $p$ variables:
\begin{equation}
\label{eq:Atxp} A_t(x,p)= (x,p \,) A(t) \Big( \begin{array}{c}
  x \\
  p \\
\end{array} \Big),
\end{equation}
with
\begin{equation}
A(t)= e^{- \Gamma(t)} \int_0^t e^{\Gamma(t_1)} R^{\dag}(t,t_1)
M(t) R(t,t_1) dt_1. \label{eq:At}
\end{equation}
In this equation, the matrix $R(t,t_1)$ contains rapidly
oscillating terms and $M(t)$ is given by
\begin{equation}
\label{eq:Mt}
\left(%
\begin{array}{cc}
  \Delta(t) & - \Pi(t)/2 \\
  - \Pi(t)/2 & 0 \\
\end{array}%
\right),
\end{equation}
with $\Delta(t)$ and $\Pi(t)$ diffusion coefficients present in
the Master Equation (\ref{QBMme}). Finally the time dependent term
$\Gamma(t)$ appearing in Eqs.~(\ref{chit}) and (\ref{eq:At}) is
simply related to the dissipation coefficient $\gamma(t)$:
\begin{equation}
\Gamma(t)= 2\int_0^t \gamma(t_1)\:dt_1. \label{Gamma}
\end{equation}
Eq.~(\ref{chit}) shows that the QCF is the product of an
exponential factor, depending on both the diffusion coefficients
[$\Delta(t)$ and $\Pi(t)$] and the dissipation coefficient
[$\gamma(t)$], and a transformed initial QCF. The exponential term
accounts for energy dissipation and is independent of the initial
state of the system. Information on the initial state is given by
the second term of the product, the transformed initial QCF. Note
that for long times $ \chi_0 \left[ e^{- \Gamma (t)} \tilde{x},
e^{- \Gamma (t)} \tilde{p} \right] \rightarrow 1$, and the system
approaches, as one would expect, a thermal state at the reservoir
temperature, whatever the initial state was.

\section{Rotating Wave Approximations}
In this section we discuss the existence of two approximations
often called with the very same name: RWA. Such a situation, of
course, may cause misleading and, in some cases, can lead to an
inaccurate description of the short time dynamics of a damped
harmonic oscillator. A similar analysis on the RWA has been
performed by Agarwal for the Master Equation describing
spontaneous emission in a two-level system~\cite{agarwal}.

Let us begin discussing what we will call  the
{\it \lq RWA performed after tracing over the environment\rq}.
Such an approximation consists in averaging over the rapidly
oscillating terms contained in the matrices $R(t,t_1)$, appearing
in Eq. (\ref{eq:At}) (for the precise analytical calculation see
Ref.~\cite{PRAsolanalitica}). This approximation can be actually
seen as a secular approximation, and therefore to avoid confusion
we will call it hereafter with this name. Note that the
microscopic interaction Hamiltonian contains the so-called
 counter--rotating
terms not conserving the unperturbed energy of the system
and reservoir. The crucial point to stress is that the
contribution of these terms is not washed out by the average over
the rapidly oscillating terms, as we will show with a specific
example later in this section.

The second type of RWA is what we will call \lq{\it RWA performed
before tracing over the environment}\rq, or simply the RWA. In
this case the counter--rotating terms are already absent in the
microscopic system--reservoir interaction Hamiltonian which reads:
\begin{equation}
H_{I}=\sum_n \left( g_n a^{\dag} b_n + h.c. \right),
\label{MEmicroRWA}
\end{equation}
with $b_n$ annihilation operators of the reservoir harmonic
oscillators, $a=\left( X+i P \right)/\sqrt{2}$ and
$a^{\dag}=\left( X-i P \right)/\sqrt{2}$. This interaction
Hamiltonian is very common in Quantum Optics, but does not give an
appropriate description of the dynamics for short times, since the
virtual processes associated to the counter--rotating terms play
an important role even for weak couplings.

To  illustrate better this point we consider the short time
dynamics of a specific observable for the system: its mean energy
$H_0= \omega_0 (n + 1/2)$, with $n$ quantum number operator. $H_0$
belongs to a class of operators which do not depend on the rapidly
oscillating terms averaged in the secular approximation \cite{PRAsolanalitica}. Hence for
calculating the mean value of these operators we can use the
simpler secular solution of the density matrix. The expression
obtained in this way is exact. It has been shown
\cite{PRAsolanalitica} that a sufficient condition to single out
this class of operators is the following
\begin{equation}
\label{condition}
\left[(\mathbf{X}^S)^2-(\mathbf{P}^S)^2\right]A=0,
\hspace{1cm} \mathbf{X}^S\mathbf{P}^SA=0,
\end{equation}
with $A$ generic observable of the system.

Let us now compare the short time expressions of the mean quantum
number $n$ of the system obtained after performing the secular and
RWA approximations. For simplicity we will
consider the case of weak coupling between system and a reservoir
at $T$ temperature. It is not difficult to prove that the
short time non--Markovian expression of $n(t)$, in the secular
approximation, can be written as
follows \cite{EPJRWA}
\begin{equation}
\langle n \rangle(t\ll\omega_c^{-1})\rightarrow \left[2\alpha^2
\int_0^{\infty} \omega |
g(\omega)|^2\left(n(\omega)+\frac{1}{2}\right)d\omega \right]
\frac{t^2}{2}, \label{nshort}
\end{equation}
where $n(\omega)$ is the mean number of reservoir excitations at
$T$ temperature and $g(\omega)$ is the reservoir spectral density
with cut-off frequency $\omega_c$. For the considerations made
earlier, this expression coincides with the exact mean value
of $n$ (see also ref.~\cite{grab}).

A similar calculation shows that, if we perform the \lq{\it RWA
before tracing over the environment}\rq, the short time
expression of the mean number $\langle n \rangle_{RWA}
(t\ll\omega_c^{-1})$ is exactly half of $\langle n
\rangle(t\ll\omega_c^{-1})$ as given by Eq. (\ref{nshort}). Such a
circumstance can be directly related to the fact that, in second
order perturbation theory, the virtual processes due to the
counter rotating terms combine to give rise to a real process.
Thus, neglecting them, as it is done in Eq.~(\ref{MEmicroRWA}),
amounts at neglecting one of the two channels through which the
system can exchange energy with the environment. For this reason
the variation of energy predicted in the  RWA
 is only half of the exact value.

\section{Non--Markovian Dynamics}
In this section we will exploit the analytic solution presented in
Sec. II to clarify some common misbelief related to non--Markovian
Master Equations and non--Markovian dynamics. In order to do this
we  consider again the specific example presented in previous
section, that is the dynamics of the mean energy of the system. In
view of the considerations made before, we can simplify the
discussion concerning the dynamics of this observable by
considering the solution of the Master Equation in the secular
approximation \cite{EPJRWA,letteranostra}:
\begin{eqnarray}
\frac{ d \rho_S}{d t}= &-& \frac{\Delta(t) + \gamma (t)}{2} \left[
a^{\dag} a \rho_S - 2 a \rho_S a^{\dag} + \rho_S a^{\dag} a
\right]
\nonumber \\
&-& \frac{\Delta(t) - \gamma (t)}{2} \left[  a a^{\dag} \rho_S - 2
a^{\dag} \rho_S a + \rho_S a a^{\dag}
 \right]. \label{MERWA}
\end{eqnarray}
As for Eq.~(\ref{QBMme}), we emphasize that this Master Equation,
even if non--Markovian, does not contain reservoir memory kernels
usually associated to non--Markovian generalized Master Equations.
In other words, Eq.~(\ref{MERWA}), as well as Eq.~(\ref{QBMme}), is
local in time. By definition this means that such an equation of
motion for the density matrix is characterized by the fact that
the time derivative of $\rho_S(t)$ only depends on the actual
value of $\rho_S(t)$. On the contrary, in non-Markovian Master
Equations involving memory kernels, the time derivative of
$\rho_S(t)$ is related to values $\rho_S(t')$ of the density
matrix at times $t' < t$. Dealing with a non--Markovian Master
Equation local in time is of course an advantage compared to a
description involving memory kernels, since the memory effects of
the environment are incorporated in the time dependence of the
coefficients. However, the locality of the Master Equations
(\ref{QBMme}) and (\ref{MERWA}) could seem hard to reconcile with
the intuitive idea of the effects that a generic environment may
produce.

As noticed in Ref.~\cite{romero}, it is the linearity of
the microscopic Hamiltonian model for QBM which forces the Master
Equation to be local in time. Indeed,  as a consequence of the
linear microscopic system--reservoir interaction Hamiltonian, the
density matrix of the reduced system is the solution of a linear
integro-differential equation (containing the memory kernel). The
key point is that a function $f(t)$ satisfying a linear
integro-differential equation, as the one satisfied by the density
matrix, does not depend on its entire history. Its future behavior
is uniquely determined by the Cauchy data, i.e., its initial
value and the initial value of its derivative. For the case of a
damped harmonic oscillator, the non--Markovian features are
restricted by linearity to the \lq memory\rq of the initial time
instant \cite{romero}. For this reason, the time dependent
coefficient provide all the memory effects in the evolution of the
density matrix. It is worth reminding that, as underlined by Paz
and Zurek in \cite{leshouches}, \lq perturbative Master
Equations can always be shown to be local in time\rq. In the
case of the damped harmonic oscillator this is true also for the
exact Master Equation \cite{leshouches}.

Another aspect worth stressing is related to the structure of the
 Master Equations (\ref{QBMme}) and its
approximated form (\ref{MERWA}). It is well known that, when the
Markovian approximation is performed, the Lindblad theorem ensures
that the Master Equation describing the system dynamics can be put
in the {\it Lindblad form}. In general, neither Eq. (\ref{QBMme})
nor Eq. (\ref{MERWA}), both non--Markovian, are (or can be recast)
in Lindblad form. A question therefore may arise naturally: should
we care about the positivity of the density matrix? The reason why
we ask this question is  related to Lindblad theorem, ensuring the
positivity of a density matrix satisfying a  Master Equation of
Lindblad form. When the Master Equation is not in the Lindblad
form, as it is in our case,
 there can be situations in which symptoms of unphysical
 approximations show up. For example, negative eigenvalues of
 the density matrix may appear. This is for instance the case of
 the Caldeira--Leggett model when some particular initial
 conditions, not consistent with the high temperature
approximation, are assumed.

In general, however, it is certainly not true that whenever a
Master Equation is not of Lindblad form then it does not preserve
the positivity of the density matrix. A basic assumption of the
theorem is indeed that the reduced system dynamics constitutes a
dynamical semigroup \cite{petruccionebook}. Thus, it can happen
that the Master Equation is not in Lindblad form but conserves the
positivity of the density matrix, while it violates the semigroup
property. This is actually our case, as one can realize from the
time dependence of the coefficients of the Master Equations
(\ref{QBMme}) and (\ref{MERWA}).

Now, let us have a closer look at the approximated Master Equation
(\ref{MERWA}).
 The form of this equation is
similar to the Lindblad form, with the only difference that the
coefficients appearing in the Master Equation  (\ref{MERWA}) are
time dependent. For the sake of brevity we will denote hereafter
this Lindblad--type Master Equation with time dependent
coefficients simply as {\it Lindblad--type} Master Equation {\it
whenever the time dependent coefficients are positive}. Note,
however,  that {\it Lindblad--type} Master Equations, contrarily
to Master Equations of {\it Lindblad form} (having constant
coefficients), do not satisfy the semigroup property. The reason
why we focus on the positivity of the time dependent coefficients
of Eq. (\ref{MERWA}) is that the dynamics of the system depends
crucially on their sign. One can define two different regions of
the parameter space, the first one correspondent to the situation
in which the Master Equation coefficients are positive (Lindblad
type region) and the second one correspondent to the case in which
the coefficients acquire, at some time instants, negative values
(non--Lindblad type region). In Ref. \cite{inprep} a detailed
study of the border between the Lindblad and non--Lindlad type
regions is carried out and the reservoir parameters governing the
passage from one region to the other one are singled out.

The first and important property distinguishing the Lindblad from
the non--Lindblad type regions  is that in the first one, i.e.
whenever the time dependent coefficients of the Master Equation
are positive, one may use the \lq standard\rq Monte Carlo wave
function method to unravel Eq.~(\ref{MERWA})~\cite{MCWF}, and
therefore to simulate the system dynamics. Such method is usually
used to study numerically the time evolution of Markovian systems
since, in order to apply it, one needs to cast the Master Equation
in the Lindblad form. Here we note that also when the Markovian
approximation is not performed, if the Master Equation is of
Lindblad type, the method can be applied.
 This is, for instance,  the case of a
damped harmonic oscillator interacting with a high temperature
reservoir, as far as the reservoir cut-off frequency  is bigger
than the frequency of the system oscillator \cite{letteranostra}.
 We note that up to recently, only quantum state diffusion unravelings were used for
simulating the temporal behavior of a harmonic oscillator
interacting with a non-Markovian quantized environment \cite{QSD}.

We now proceed to illustrate an example showing the short time
dynamics of the system in the non--Lindblad type region, that is
 when the time dependent coefficients of the Master
Equation are negative. To this aim we consider the case of the
mean energy of the system oscillator when it interacts with an
artificially engineered out of resonance reservoir. In order to
study this example we first write down the density matrix solution
in the secular approximation.

This approximation washes out the contribution of the diffusive
term $\Pi (t)$ in Eq.~(\ref{eq:At}), which becomes
\cite{PRAsolanalitica}
\begin{equation}
A(t)= e^{- \Gamma(t)} \int_0^t e^{\Gamma(t_1)} \Delta(t_1) dt_1
\equiv \Delta_{\Gamma}(t). \label{eq:Atappr}
\end{equation}
The simplified QCF thus reads
\begin{equation}
\chi_t (x,p)=e^{- \Delta_{\Gamma}(t) (x^2+p^2)/2} \chi_0 \left[
e^{- \Gamma (t)} \tilde{x}, e^{- \Gamma (t)} \tilde{p}  \right].
\label{chit2}
\end{equation}
By using Eq.~(\ref{xp}) one derives the following expression for
the mean energy of the system \cite{PRAsolanalitica}
\begin{equation}
\langle H_0 (t) \rangle =e^{-\Gamma(t)} \langle H_0 (t=0) \rangle
+ \omega_0 \Delta_{\Gamma}(t). \label{eq:H0media}
\end{equation}

Let us now consider the case in which the system oscillator is in
its ground state, and the environment is in a thermal state at $T$
temperature. We assume a Ohmic reservoir spectral density with
Lorentz-Drude cut--off $\omega_c$. Moreover, we consider the case
in which $\omega_c < \omega_0$, that is the spectrum of the
reservoir do not completely overlap  with the frequency of the
system oscillator. This is of course never the case in presence of
a natural reservoir. However, it has been recently shown that for
quasi--closed systems, as single trapped ions, it is possible to
engineer different types of artificial reservoirs, and couple them
in a controlled way to the ion motion~\cite{engineerNIST}.

\begin{figure}
\includegraphics[width=8 cm,height=5cm]{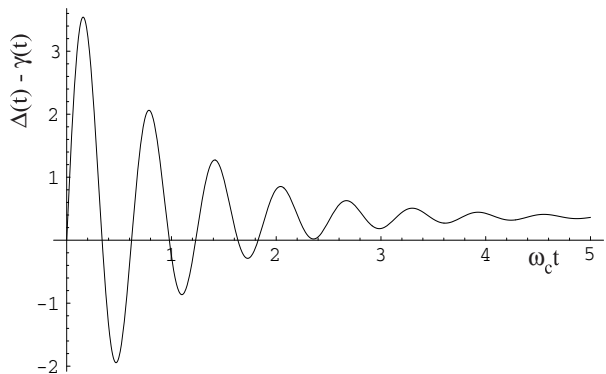}
\caption{\label{fig1} \footnotesize Short time behavior of the
 time dependent coefficient $\Delta(t)- \gamma(t)$ for $\alpha \omega_0= 0.1$Hz,
 $\omega_0= 10^7$Hz, $r=0.1$, and $T=300$K.}
\end{figure}

\begin{figure}
\includegraphics[width=8 cm,height=7cm]{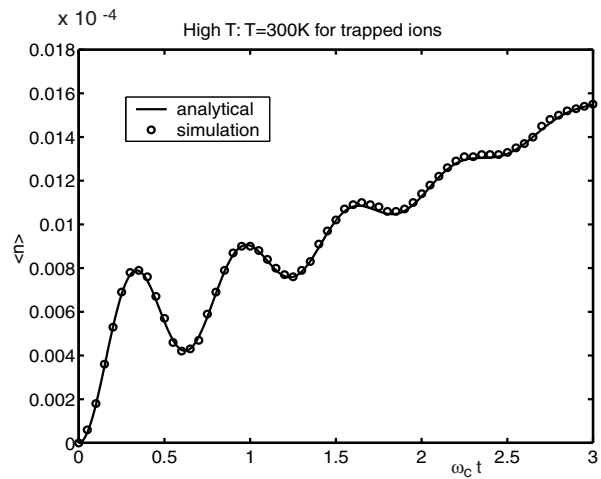}
\caption{\label{fig2} \footnotesize Short time behavior of the
heating function
 for dimensionless coupling constant  $\alpha \omega_0= 0.1$Hz,
 $\omega_0= 10^7$Hz, $r=0.1$, and $T=300$K. We
compare the analytic solution and the Monte Carlo simulation with
$10^7$ histories.}
\end{figure}

Under the conditions stated above, for high reservoir temperature,
the time dependent coefficient $\Delta(t)- \gamma(t)$ shows an
oscillatory behavior as depicted in Fig.~\ref{fig1}. As one can
see from this figure there exist intervals of time in
correspondence of which $\Delta(t)-\gamma(t)$ becomes negative.
Hence, in general, the Master Equation (\ref{MERWA}) is not of
{\it Lindblad--type}. The density matrix, however, is always
positive and it is the semigroup property which is clearly
violated, as one can see from Fig.~\ref{fig2} where the time
evolution of the mean quantum number operator $\langle n\rangle$
(heating function) is plotted. The signature of the non--Lindblad
type dynamics of the reduced system is given by the presence of
oscillations in its mean energy. These oscillations are due to
virtual exchanges of energy between the system and the reservoir.
As we have shown in references~\cite{EPJRWA} and ~\cite{inprep},
these virtual processes are absent whenever the Master Equation
(\ref{MERWA}) is of Lindblad--type.
In Figure \ref{fig2} the analytic result given by Eq.
(\ref{eq:H0media}) is compared with the simulation using the
Non--Markovian Monte Carlo method
\cite{Breuer99a,petruccionebook}. In this case indeed, as we have
previously discussed, the \lq standard \rq Monte Carlo wave
function method cannot be applied.

\section{Conclusions}\label{sec:conclusion}
In this paper we have used the exact solution of a damped harmonic
oscillator \cite{PRAsolanalitica} to discuss some relevant issues
of open system dynamics often leading to confusion and/or
misunderstandings.

We have focussed our attention on the so called RWA and we have
pointed out the existence of two different approximations called
with the very same name. In other words we have stressed the
differences in what we call the \lq{\it RWA performed before or after
tracing over the environment}\rq. The {RWA performed before tracing
over the environment} consists in neglecting the counter--rotating
terms in the microscopic Hamiltonian describing the coupling
between system and environment. The {RWA performed after tracing
over the environment} is more precisely a secular approximation,
consisting in an average over rapidly oscillating terms, but does
not wash out the effect of the counter--rotating terms present in
the coupling Hamiltonian. By considering a specific example we
show how, in the short time non--Markovian regime, the virtual
processes due to the counter--rotating terms (neglected in the RWA)
give a significant contribution even for weak
coupling, and thus need to be taken into account.

We have pointed out that non-Markovian Master Equations do not
necessarily contain memory kernels. This concept, already claimed
by many authors (see for example
\cite{petruccionebook,leshouches}) is still - surprisingly - often
seen in a rather sceptical way. We have also discussed a related
point which often rises doubts: the conservation of the positivity
of the density matrix when the Master Equation ruling its dynamics
is not in Lindblad form.

In order to study the dynamical properties of non--Markovian
Master Equations, we have introduced two subclasses of Master
Equations which are not in the Lindblad form: the Lindblad and
non--Lindblad type, and we have stressed their difference. We have
presented a new example showing that there exist conditions under
which the short time dynamics of a damped oscillator is governed
by a non--Lindblad type Master Equation whose solution is always
positive. In this case  the semigroup property for the reduced
dynamics which is violated, and the non Lindblad--type dynamics
shows up in the existence of virtual energy exchanges between
system and reservoir.

We think that this paper contributes in clarifying
some aspects, often mislead, of the non--Markovian dynamics of a damped harmonic.
This is important also because very
recently the potential interest in non--Markovian reservoirs for
quantum information processing has been demonstrated \cite{Ahn}
and a non--Markovian description of quantum computing, showing the
limits of the Markovian approach, has been presented \cite{horo}.

\section{Acknowledgements}
J.P. acknowledges financial support from the Academy of Finland
(project no.~50314), the Finnish IT-Center for Science (CSC) for
the computer resources, the University of Palermo for the
hospitality, and Kalle-Antti Suominen
for the discussions.
S.M. acknowledges Finanziamento Progetto Giovani
Ricercatori anno 1999, Comitato 02 for financial support.

\end{document}